\documentclass[useAMS,usenatbib,referee]{biom}
\usepackage{color}

\def\bSig\mathbf{\Sigma}

\newcommand{\bea}{\begin{eqnarray*}}
\newcommand{\eea}{\end{eqnarray*}}
\newcommand{\ba}{\begin{array}}
\newcommand{\ea}{\end{array}}
\newcommand{\be}{\begin{enumerate}}
\newcommand{\ee}{\end{enumerate}}

\usepackage[figuresright]{rotating}
\usepackage{hyperref}
\usepackage{mathtools}

\title[Enriched Regression]{On Data Enriched Logistic Regression}

\author{Cheng Zheng$^{1}$, 
Sayan Dasgupta$^{2}$, Yuxiang Xie$^{3}$, Asad Haris$^{3}$ and Ying Qing Chen$^{2,*}$\email{yqchen@fhcrc.org} \\
$^{1}$Joseph. J. Zilber School of Public Health, University of Wisconsin-Milwaukee, Milwaukee, USA \\
$^{2}$Biostatistics and Biomathematics Program, Fred Hutchinson Cancer Research Center, Seattle, USA \\
$^{3}$Department of Biostatistics, University of Washington, Seattle, USA}
\begin{document}


\pagerange{\pageref{firstpage}--\pageref{lastpage}} \pubyear{2018}

\volume{}
\artmonth{}
\doi{}


\label{firstpage}


\begin{abstract}
Biomedical researchers usually study the effects of certain exposures on disease risks among a well-defined population. To achieve this goal, the gold standard is to design a trial with an appropriate sample from that population. Due to the high cost of such trials, usually the sample size collected is limited and is not enough to accurately estimate some exposures' effect. In this paper, we discuss how to leverage the information from external `big data' (data with much larger sample size) to improve the estimation  accuracy at the risk of introducing small bias. We proposed a family of weighted estimators to balance the bias increase and variance reduction when including the big data. We connect our proposed estimator to the established penalized regression estimators.  We derive the optimal weights using both second order and higher order asymptotic expansions. Using extensive simulation studies, we showed that the improvement in terms of mean square error (MSE) for the regression coefficient can be substantial even with finite sample sizes and our weighted method outperformed the existing methods such as penalized regression and James Stein's approach. Also we provide theoretical guarantee that the proposed estimators will never lead to asymptotic MSE larger than the maximum likelihood estimator using small data only in general. We applied our proposed methods to the Asia Cohort Consortium China cohort data to estimate the relationships between age, BMI, smoking, alcohol use and mortality.  
\end{abstract}

%
%


\maketitle
\section{Introduction}
In most research settings in medicine, we aim at knowing the effect of a certain exposure to the risk of some specific disease among a well-defined targeted population. To achieve this goal, well-designed trials are usually used. However, the sample sizes for such studies are usually limited due to the high cost of recruitment and thus the sample size usually just have power to detect the effect for the primary exposure of interest. In the mean time, the number of available observational studies or trial studies from other population are accumulated quickly nowadays. Can we use information from these data to improve our inference on the population where small data is drawn from? Here we refer the randomized clinical trial data which have a clearly defined target population by the design and sampling scheme as ``small data'' and refer other external data as ``big data''. Our goal is to efficiently combine the information from these two types of data to obtain more accurate estimation of the association between some readily available quantities that are presented in both data (e.g., age, gender) and risk of diseases among population where small data is drawn from. Since the distributions of predictors as wells as the relationship among these predictors and the event of interest are likely to be different between ``big data'' and ``small data'', we are at the risk of introducing some bias when using the information from big data. However, given the size of the big data, they can still provide insightful information about how we predict risk among the targeted population, assuming that the two populations share a certain degree of similarity in their prediction model forms. This motivate us to find an estimator that is better than using ``small data'' only  with a bias-variance trade-off. Directly pooling the two data together can lead to substantial bias and lead to increased mean square error (MSE) when the difference between the two sources are large. This motivate us to find estimators that are always not lead to increased MSE than using ``small data'' only and do lead to decreased MSE under certain situation.

Previous studies have shown the plausibility of this type of idea. In the simple mean estimation, \cite{stein1981} showed that the simple sample mean is inadmissible. \cite{chen2015} studied the combining of regression results from small and big data in linear regression setting and showed Stein-type result for Gaussian responses: i.e., the use of small data only is inadmissible when $p\geq 5$ and degree of freedom is more than 10. \cite{gross2016} proposed to use shared Lasso to achieve this in linear regression setting. However, the similar enhanced regression approach for non-Gaussian outcome has not been fully studied. To our knowledge, the risk prediction work are mostly dependent on the assumption that certain reduced marginal model or marginal information from the big data is accurate \citep{cheng2018, chatterjee2016}. In this work, we propose to fill the gap of risk prediction combining information from small and big data for binary outcome that relies on an alternative structural assumption where we assume the effect structure rather than effect magnitude are the same.

The structure of the paper is as follow. In section 2, we introduce the notation and models we used followed by our proposed estimators and its connection to other existing estimators. In section 3, we study the performance of different estimators for their finite sample properties and show the improvement using our proposed estimator. In section 4, we provide theoretical results that guarantee our proposed estimator to be no worse than the small data only analysis in terms of MSE. In section 5, we applied our method to analyze the Asia Cohort Consortium data with sensitivity for potential violation of model assumptions. In section 6, we discuss the potential extension of our proposed estimator to more general setting.

\section{Methods}
\subsection{Notation and Model}
We denote our outcome of interest as $Y\in \{0,1\}^{n_S+n_B}$ and denote the design matrices by $X_S\in R^{n_S\times p}$ and $X_B\in R^{n_B\times p}$ where $n_S$ and $n_B$ represent the sample sizes for the small data and the big data respectively. In general, we reserve the subscripts $B$ and $S$ to denote quantities related to the big data and the small data respectively. Since the outcomes of interest is binary (disease occurrence), denoting $\mu_i=E(Y_i|X_i)=Pr(Y_i|X_i)$, we assume logistic regression models for both the small data and the large data and write them as: 
\begin{eqnarray}
\label{eqmod}
&&\log \left(\frac{\mu_i}{1-\mu_i}\right)=X_i\left\{\beta+\gamma I(i\in B)\right\}
\end{eqnarray} 
where $\beta\in R^{p\times 1}$ and $\gamma\in R^{p\times 1}$ are unknown regression parameters and $I(\cdot)$ is indicator function. Our goal is to obtain accurate estimation of $\beta$ while treating $\gamma$ as nuisance parameter using the information from both the small and the big data. In this project, we propose novel weighted shrinkage estimators and relate them to the penalized regression based estimators. We compare the performance of weighted estimators, penalized estimators and the James-Stein type shrinkage estimator \citep{James1961,Efron1973}.


\subsection{Penalized regression based estimators}
Here we first introduce the penalized regression based estimators that can be used to integrate the information from the two data sets. We consider minimizing the following object function in order to obtain an estimates of $\beta$, where a penalty is put on $\gamma$ only.
\begin{eqnarray*}
L_{\lambda}(\beta,\gamma)&=&-\sum_{i=1}^{n_S+n_B}\{y_i\log\frac{\mu_i}{1-\mu_i}+\log (1-\mu_i)\}+p_{\lambda}(\gamma)\\
&=&-\sum_{i=1}^{n_S+n_B} y_ix_i\left\{\beta+\gamma I(i\in B)\right\}-\log \left\{1+\exp\{\beta+\gamma I(i\in B)\}\right\}+p_{\lambda}(\gamma)
\end{eqnarray*} 
where $p_{\lambda}(\gamma)$ is the penalty term. It is obvious that the estimators from above optimization problem can be implemented using a penalized logistic regression such as \textit{glmnet} in R \citep{Friedman2008} with design matrix that has row like $(x_i,x_iI(i\in B))$ and the penalty factor $(0_p,1_p)$. Here the form of the penalty term can be flexible, for example, $p_{\lambda}(\gamma)=\lambda\sum_{j=1}^p|\gamma_j|$ for LASSO $L_1$ penalty \citep{tibshirani1996,efron2004}, $p_{\lambda}(\gamma)=\lambda\sum_{j=1}^p\gamma^2_j$ for ridge regression $L_2$ penalty. Other penalties like elastic net \citep{zou2005}, SCAD \citep{fan2001}, MCP \citep{zhang2010} can also be used but for comparison purpose, we use $L_1$ and $L_2$ penalty to represent the performance of this class of estimators.

When the prediction in small dataset is more important than the estimation of regression parameter $\beta$, instead of penalize based on parameter, we might penalize on the extra linear predictor $X_s\gamma$ and use $p_{\lambda}(X_s\gamma)$ to replace $p_{\lambda}(\gamma)$. The tuning parameter $\lambda$ could be determined via $K-$fold cross-validation \cite{arlot2010} in small data set. 



\subsection{Weighted shrinkage estimator}
We propose an alternative approach to the penalized regression method via weighted shrinkage method. This method has been shown to be useful under linear regression model \citep{chen2015}, however, as we will see in this section, the application of it to this nonlinear model is not straightforward.

The basic idea of this kind of the weighted shrinkage estimator is to first fit the logistic regression model among the small data and the big data separately to obtain $\hat{\beta}_S$ and $\hat{\beta}_B$ as the estimator for $\beta$ and $\beta+\gamma$ and then combine the two estimators through an weighted average $\hat{\beta}_W=W\hat{\beta}_S+(I-W)\hat{\beta}_B$ for a specific weight matrix $W\in R^{p\times p}$. Specifically, when $W=I$, this is just the estimator of using small data only and when $W=\frac{n_S}{n_S+n_B}I$, this can be approximately viewed as a pooled estimator of the small and the big dataset assuming $\gamma=0$. 

It is obvious that the performance of $\hat{\beta}_W$ highly depend on the choice of weight matrix $W$. The major goal here is to find the optimal weight matrix as a function of $\beta$, $\gamma$ and data $X_S$, $X_B$ where $X_S\in R^{n_S\times p}$ and $X_B\in R^{n_B\times p}$ are design matrices for small and large data respectively. Here we define optimal weight by the weight that minimize the coefficient estimation error $E||\hat{\beta}_W-\beta||_2^2$.


To find out the form for the optimal weight, we use the asymptotic expansion of $\hat{\beta}_S$ and $\hat{\beta}_B$. The optimal weight obtained via second-order approximation is denoted as $W_2(\beta,\gamma,X_S,X_B)$ and the optimal weight obtained via higher-order Edgeworth expansion \citep{hall1992} is denoted as  $W_h(\beta,\gamma,X_S,X_B)$. For all these weights, we could plug in $\hat{\beta}_S$ and $\hat{\beta}_B-\hat{\beta}_S$ for $\beta$ and $\gamma$ to obtain estimated version of these optimal weights.



For the James-Stein estimator, we use the form from \cite{an2010}, i.e, $W_{JS}=diag\{(1-c/F_i)\}$ or $W_{JS+}=diag\{(1-c/F_i)_{+}\}$ where $x_{+}=\max\{0,x\}$ and $F_i$, $i=1,\cdots,p$ is the test statistic for $\hat{\beta}_{Si}=\hat{\beta}_{Bi}$. Here we have $c=\frac{(p-2)(n-2)}{pn}$.

Now we provide more details on how to obtain these weights. We begin with the second order approximation. Using the expansion of logistic regression estimator, we have
\begin{eqnarray*}
\hat{\beta}_S&=&\beta+(X_S^TV_SX_S)^{-1}X_S^T(Y_S-\mu_S)+O_p(n_S^{-1})\\
\hat{\beta}_B&=&\beta+\gamma+(X_B^TV_BX_B)^{-1}X_B^T(Y_B-\mu_B)+O_p(n_B^{-1}).
\end{eqnarray*}
where $V_{S}=\mu_S(1-\mu_S)$ are the variance of $Y_S$ and $V_{B}=\mu_B(1-\mu_B)$ are the variance of $Y_B$. Ignoring $O_p(n_S^{-1})$ and $O_p(n_B^{-1})$ terms in the above expansion lead to second order optimal weight 
\begin{equation}
\label{eqW2}
W_2(\beta,\gamma,X_S,X_B)=(\gamma\gamma^T+\Sigma_B^{-1}+\Sigma_S^{-1})^{-1}(\gamma\gamma^T+\Sigma_B^{-1}).
\end{equation}
where $\Sigma_S=X_S^TV_SX_S$, $\Sigma_B=X_B^TV_BX_B$ and corresponding estimated second order optimal weight $\hat{W}_2(\hat{\beta},\hat{\gamma},X_S,X_B)$ where $\hat{W}_2$ means the terms $\Sigma_S$ and $\Sigma_B$ are replaced by their consistent estimated version.

For the higher order approximation, we use the approximation 
\begin{eqnarray}
\hat{\beta}_S&=&\beta+n_S^{-1/2}B_S+n_S^{-1}C_S+n_S^{-3/2}D_S+n_S^{-2}E_S+o_p(n_S^{-2})\\
\hat{\beta}_B&=&\beta+\gamma+n_B^{-1/2}B_B+n_B^{-1}C_B+n_B^{-3/2}D_B+n_B^{-2}E_B+o_p(n_B^{-2})
\end{eqnarray}
where $B_S=n_S^{1/2}(X_S^TV_SX_S)^{-1}X_S^T(Y_S-\mu_S)$, $B_B=n_B^{1/2}(X_B^TV_BX_B)^{-1}X_B^T(Y_B-\mu_B)$ and the expression of higher order terms could be found in the appendix. We have $EB_S=0$ and $Var(B_S)=n_S\Sigma_S^{-1}$,$EB_B=0$ and $Var(B_B)=n_B\Sigma_B^{-1}$. Denote $c_S=EC_S$, $v_S=Var(C_S)$,  $d_S=ED_S$,  $e_S=EE_S$ and $\rho_S=Cov(B_S,C_S)=<B_S,C_S>$, $\nu_S=Cov(B_S,D_S)=<B_S,D_S>$, $c_B=EC_B$, $v_B=Var(C_B)$, $d_B=ED_B$, $e_B=EE_B$ and $\rho_B=Cov(B_B,C_B)=<B_B,C_B>$, $\nu_B=Cov(B_B,D_B)=<B_B,D_B>$. Ignore $o_p(n_S^{-2})$, $o_p(n_B^{-2})$ terms, we have $E||\hat{\beta}_W-\beta||_2^2$
is minimized at 
\begin{equation}
\label{eqWh}
W_h(\beta,\gamma,X_S,X_B)=(A_{11}-A_{10})(A_{00}+A_{11}-A_{01}-A_{10})^{-1}
\end{equation}
where
\bea
A_{00}&=&\Sigma_S^{-1}+n_S^{-2}(c_Sc_S^T+v_S)+n_S^{-3/2}\rho_S+n_S^{-3/2}\rho_S^T+n_S^{-2}\nu_S+n_S^{-2}\nu_S^T\\
A_{01}&=&n_S^{-1} c_S\gamma^T+n_S^{-3/2} d_S\gamma^T+n_S^{-2}e_S\gamma^T+n_S^{-1}n_B^{-1}c_Sc_B^T\\
A_{10}&=&n_S^{-1}\gamma c_S^T+n_S^{-3/2}\gamma d_S^T +n_S^{-2}\gamma e_S^T+n_S^{-1}n_B^{-1}c_Bc_S^T\\
A_{11}&=&\gamma\gamma^T+\Sigma_B^{-1}+n_B^{-2}(c_Bc_B^T+v_B)+n_B^{-1}\gamma c_B^T+n_B^{-1}c_B\gamma^T+n_B^{-3/2}\gamma d_B^T+n_B^{-3/2}d_B\gamma^T\\
&&+n_B^{-2}\gamma e_B^T+n_B^{-2}e_B\gamma^T+n_B^{-3/2}\rho_B+n_B^{-3/2}\rho_B^T+n_B^{-2}\nu_B+n_B^{-2}\nu_B^T.\\ 
\eea
The estimated version can be denoted as $\hat{W}_h(\hat{\beta},\hat{\gamma},X_S,X_B)$ where $\hat{W}_h$ means $c_S$, $c_B$, $d_S$, $d_B$, $e_S$, $e_B$, $v_S$, $v_B$, $\rho_S$, $\rho_B$, $\nu_S$, $\nu_B$ in $W_h$ are replaced by their consistent estimators.\\

\subsection{Relationship between two types of estimators}
The shrinkage estimator defined above is closely related to the penalized estimator. $W$ can be written as $W_{\lambda}$ based on the $L_2$ penalty using the asymptotic linear expansion of GLM as in equation \ref{eqW2} and equation \ref{eqWh} are $O_p(1)$ term whose form can be obtained from Edgeworth expansion \citep{sun2000}. To relate this to the penalized regression method, we consider the following penalized version and find $\hat{\beta}_{\lambda}$ and $\hat{\gamma}_{\lambda}$ minimize
\begin{eqnarray*}
||V_S^{-1/2}(Y_S-\mu_S)||_2^2+||V_B^{-1/2}(Y_B-\mu_B)||_2^2+||\lambda^{1/2}X_T\gamma||_2^2
\end{eqnarray*}
Denote $\mathcal{X}=\left(\begin{array}{cc}X_S&0\\ X_B& X_B\\ 0 &X_T\end{array}\right)$ and $\mathcal{Y}=\left(\begin{array}{c}Y_S-\mu_S\\ Y_B-\mu_B\\ 0 \end{array}\right)$ and $\mathcal{V}=\left(\begin{array}{ccc}V_S&0&0\\ 0&V_B& 0\\ 0&0 &\lambda I\end{array}\right)$
The score function will be
\begin{eqnarray*}
U=\mathcal{X}^T\mathcal{Y}
\end{eqnarray*}
and the information matrix
\begin{eqnarray*}
\mathcal{I}=\mathcal{X}^T\mathcal{V}\mathcal{X}
\end{eqnarray*}
So we have the approximation
\begin{eqnarray*}
\left(\begin{array}{c}\hat{\beta}_{\lambda}\\ \hat{\gamma}_{\lambda}\end{array}\right)&=&\left(\begin{array}{c}\beta\\ \gamma\end{array}\right)+(\mathcal{X}^T\mathcal{V}\mathcal{X})^{-1}\mathcal{X}^T\mathcal{Y}+O_p(n_S^{-1}).\\
\end{eqnarray*}
Define $\Sigma_T=X_T^TX_T$ and 
$$W_{\lambda}=(\Sigma_S+\lambda\Sigma_T+\lambda\Sigma_T\Sigma_B^{-1}\Sigma_S)^{-1}(\Sigma_S+\lambda\Sigma_T\Sigma_B^{-1}\Sigma_S).$$ Then we have
\begin{eqnarray*}
\hat{\beta}_{\lambda}=W_{\lambda}\hat{\beta}_S+(I-W_{\lambda})\hat{\beta}_B+O_p(n_S^{-1})
\end{eqnarray*}
So we can see that with different choice of $X_T$ and $\lambda$, $L_2$ penalized estimator is asymptotically equivalent to weighted estimator.

\section{Simulations}
To see how much efficiency we can gain using our proposed estimator under finite sample setting, We use a detailed simulation exercise, as described below, to compare the different methods with each other. In these simulations, we consider different size for $p$, the dimension for $\beta$ (including intercept), such that $p\in \{3,6,11\}$, but we set the $L_2$ norm of $\beta$, i.e., $||\beta||_2$ to be fixed at $p\log(1.1)$. We also vary $\gamma$, the amount of bias in the big data, such that $||\gamma||_2/||\beta||_2$ takes the following values $\{0.5,1,2\}$. We generate the small data $X_S$ and the big data $X_B$ from the same Gaussian distribution, and the covariates are assumed to be uncorrelated with each other. We also vary $n_S$, the size of the small data $X_S$, between $100$ and $500$ in increments of $50$ (thus $n_S\in\{100,150,200,250,300,350,400,450,500\}$), while we consider 
two fixed sizes for $n_B$, namely $\{1000,10000\}$. For each simulation, we generate $Y$ based on our assumed logistic models for the small and the big data, given in equation \ref{eqmod},

For each simulation scenario, we perform 100 simulations to compute the mean squared error (MSE) in estimation, $E||\hat{\beta}-\beta||_2^2$. We obtain estimates for $\beta$ by (1) using small data only (Small), (2) pooling big and small data (Pool), (3) weighted with optimal weight from second order approximation ($W_2$), (4) weighted with optimal weight from higher order approximation ($W_h$), (5) $L_1$ penalized regression ($L_1$), (6) $L_2$ penalized regression ($L_2$), or (7) JS+ weighted estimator (JSP).

\begin{figure}[ht!]
\begin{center}
\includegraphics[width=1\textwidth]{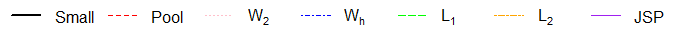} 
\includegraphics[width=1\textwidth]{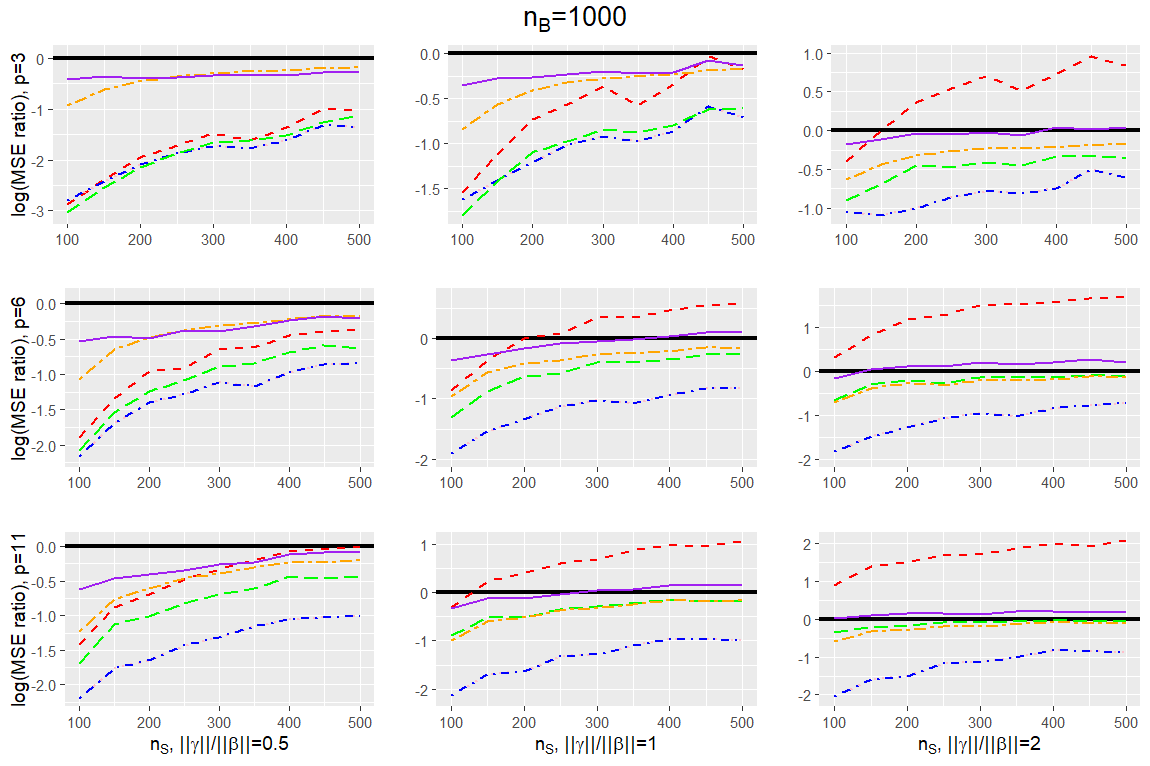} 
\caption{Plot for the log transformed ratios of the mean squared error of $\hat{\beta}$, when we use each of the procedures (Small, Pool, $W_2$, $W_h$, $L_1$, $L_2$, JSP), versus that when we only use the small data (Small), for varying sizes of $n_S$ and when $n_B$ is 1000.} 
\label{fig:nb1000}
\end{center}
\end{figure}

The simulation results are shown in Figures \ref{fig:nb1000}
and \ref{fig:nb10000}. Each figure represents the different simulation settings under a fixed size of the big data, for example, Figure \ref{fig:nb1000} shows plots for different simulation settings when $n_B=1000$, 
and Figure \ref{fig:nb10000} shows the same for $n_B=10000$. In each figure, nine plots are presented in a grid of three rows and three columns, where the columns shows plots for a particular ratio of $||\gamma||_2/||\beta||_2$ (in increasing order of magnitude from left to right), while the rows show plots for different dimension size $p$ (in increasing order of magnitude from top to bottom). Each plot in the grid presents the graphs of the log transformed ratio of the MSE of $\hat{\beta}$, when we use each of the procedures (Small, Pool, $W_2$, $W_h$, $L_1$, $L_2$, JSP), versus that when we only use the small data (Small), as a function of the varying sizes for $n_S$. From the plots, we can make the following observations:
\begin{itemize}
\item The performance of $W_2$ and $W_h$ procedures are very close to each other in every setting, pointing to the fact that probably the optimal second order approximation weights $W_2$ suffices for our problem (in fact it is difficult to visually observe the graph for $W_2$ as it is exactly overlaid by the graph for $W_h$).  
\item In every simulation setting, the $W_2$ and the $W_h$ procedures outperform every other method. The gain in performance of $W_2$ and $W_h$ over the next best performing method increases with increasing dimension size $p$, and increasing ratio of $||\gamma||_2/||\beta||_2$. The same trend is observed for both sizes for $n_B$ (1000/10000).
\item The $L_1$ penalized procedure is the third best performing method overall (after $W_2$ and $W_h$), and its performance is similar to the $W_2$/$W_h$ procedures when dimension size is small ($p=3$), and the relative bias is low ($||\gamma||_2/||\beta||_2=0.5$).  
\item The pooled procedure is the worst performing method overall and is quite sensitive to the bias $||\gamma||_2$. Although it shows relatively good performance when dimension size is small ($p=3$), and when the relative bias is low ($||\gamma||_2/||\beta||_2=0.5$), with increasing dimensions, and especially with increasing bias, its performance becomes very poor. Apart from JSP in some scenarios, it is the only procedure that show extremely elevated MSEs in comparison to the small data.
\item The performance of the $L_2$ procedure is similar to JSP in some settings, but is better than it in others. For example, with increasing dimensions, and with increasing bias, the JSP procedure sometimes tend to have higher MSE than those obtained from the small data (Small), especially when the size of the small data is on the higher end, but the $L_2$ procedure always perform better than Small.
\item All methods (except for Pool and in some instances JSP) show lower MSE than the estimates obtained from the small data itself, and the gain in efficiency is most pronounced when the size of the small data is small.
\end{itemize}




\begin{figure}[ht!]
\begin{center}
\includegraphics[width=1\textwidth]{legend.png} 
\includegraphics[width=1\textwidth]{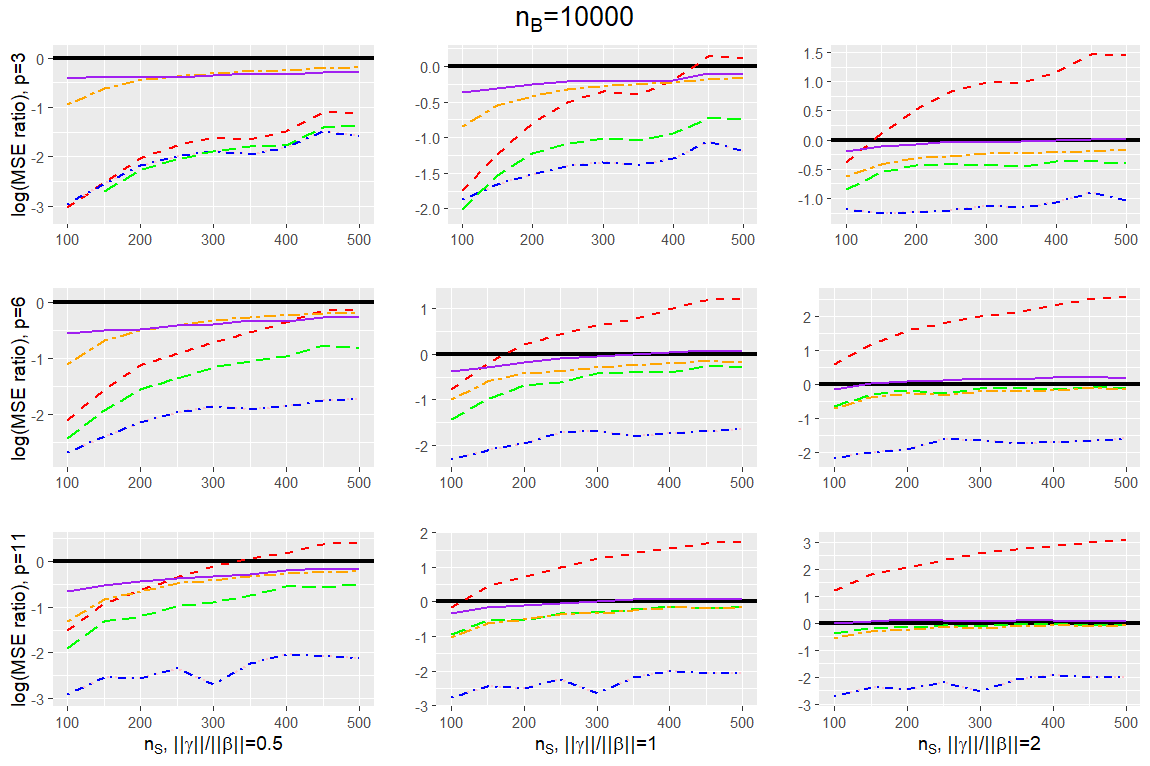}
\caption{Plot for the log transformed ratios of the mean squared error of $\hat{\beta}$, when we use each of the procedures (Small, Pool, $W_2$, $W_h$, $L_1$, $L_2$, JSP), versus that when we only use the small data (Small), for varying sizes of $n_S$ and when $n_B$ is 10000.} 
\label{fig:nb10000}
\end{center}
\end{figure}


\begin{figure}[ht!]
\begin{center}
\includegraphics[width=1\textwidth]{legend.png} 
\includegraphics[width=1\textwidth]{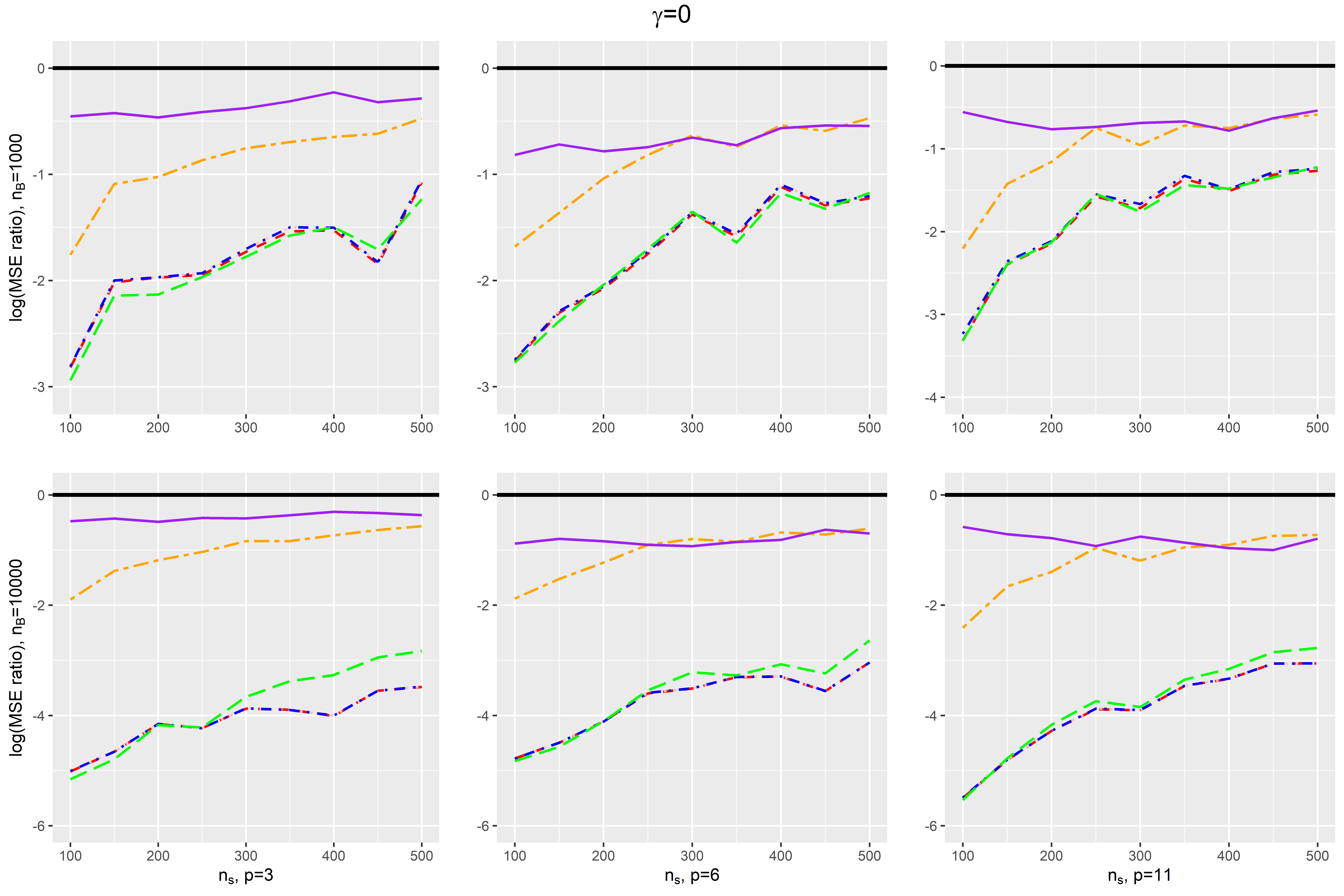}
\caption{Plot for the log transformed ratios of the mean squared error of $\hat{\beta}$, when we use each of the procedures (Small, Pool, $W_2$, $W_h$, $L_1$, $L_2$, JSP), versus that when we only use the small data (Small), for varying sizes of $n_S$ and when $\gamma=0$.} 
\label{fig:nb10000}
\end{center}
\end{figure}

\begin{figure}[ht!]
\begin{center}
\includegraphics[width=1\textwidth]{legend.png} 
\includegraphics[width=1\textwidth]{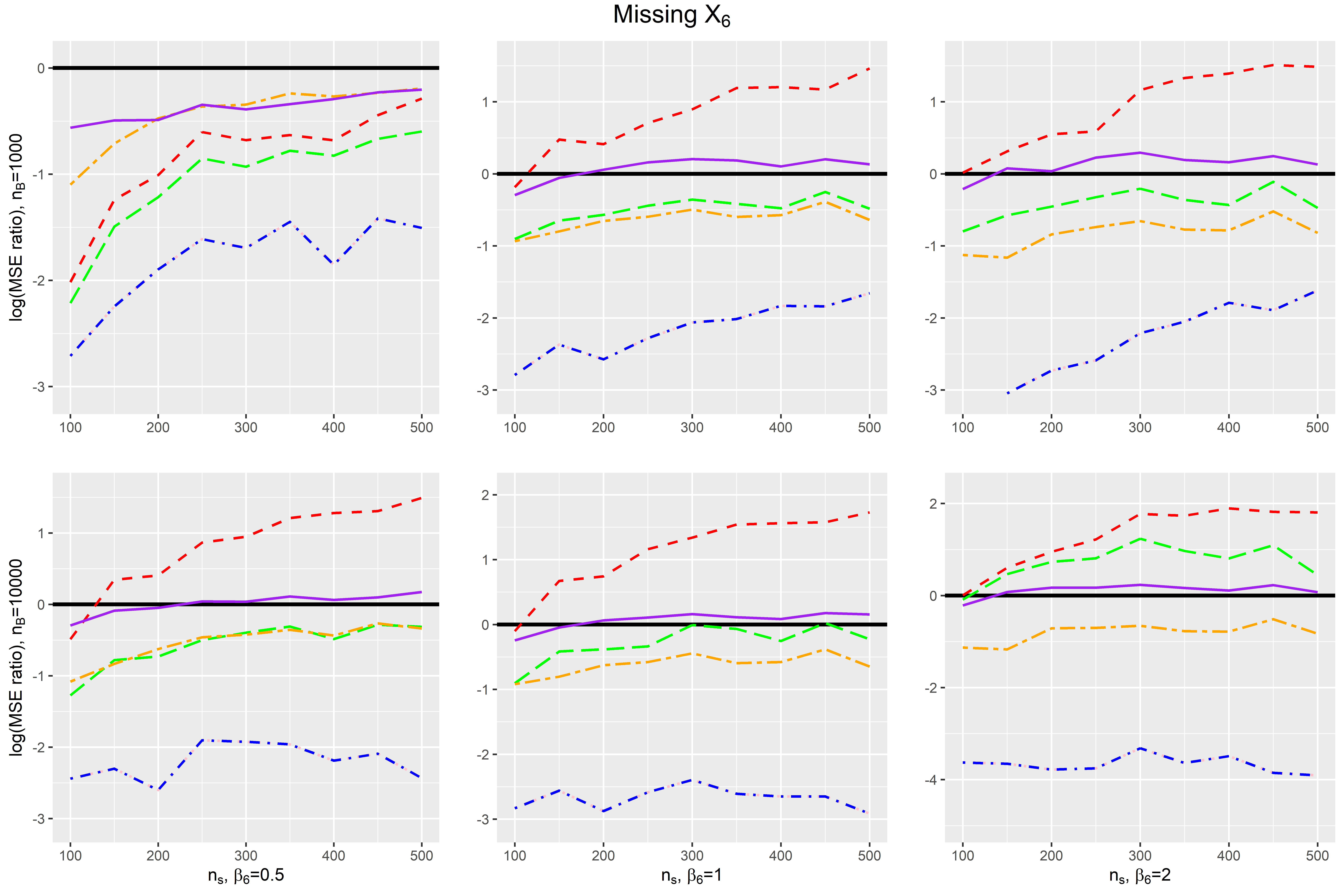}
\caption{Plot for the log transformed ratios of the mean squared error of $\hat{\beta}$, when we use each of the procedures (Small, Pool, $W_2$, $W_h$, $L_1$, $L_2$, JSP), versus that when we only use the small data (Small), for varying sizes of $n_S$ and when the bias term $\gamma$ is caused by missing the covariate $X_6$ in big data.} 
\label{fig:nb10000}
\end{center}
\end{figure}

\section{Theoretical Results}
The simulation results indicate that our proposed estimator always outperform the small-data only analysis in terms of MSE and such improvement is substantial sometimes. However, to apply our proposed method in general, a natural question is whether there exists certain scenario that the proposed estimator will underperform the small-data only analysis, especially when the difference between the two sources of data is large. To answer this question, here we summarize the theoretical guarantee of our proposed weighted estimators in the following theorems and the detail expression and proof can be found in the appendix.\\

\textbf{Theorem 1:} The second order optimal weight $W_2(\beta,\gamma,X_S,X_B)$ and its estimated version $W_2(\hat{\beta},\hat{\gamma},X_S,X_B)$ approximately minimize $E||\hat{\beta}_W-\beta||_2^2$ at $O(n_S^{-1})$ level in the sense that 
\begin{eqnarray*}
&E||\hat{\beta}_{W_2(\beta,\gamma,X_S,X_B)}-\beta||_2^2\leq \inf_W E||\hat{\beta}_{W}-\beta||_2^2+O(n_S^{-1})\\
&E||\hat{\beta}_{\hat{W}_2(\hat{\beta},\hat{\gamma},X_S,X_B)}-\beta||_2^2\leq \inf_W E||\hat{\beta}_{W}-\beta||_2^2+O(n_S^{-1}),
\end{eqnarray*}
where the infimum is taken over all $p\times p$ random matrix that are measurable given $X_S,X_B,Y$.\\
$W_2(\beta,\gamma,X_S,X_B)$ and $\hat{W}_2(\hat{\beta},\hat{\gamma},X_S,X_B)$ are also approximate optimal weight for prediction purpose in the sense that
\begin{eqnarray*}
&n_S^{-1}E||X_S(\hat{\beta}_{W_2(\beta,\gamma,X_S,X_B)}-\beta)||_2^2\leq \inf_W n_S^{-1}E||X_S(\hat{\beta}_{W}-\beta)||_2^2+O(n_S^{-1})\\
&n_S^{-1}E||X_S(\hat{\beta}_{\hat{W}_2(\hat{\beta},\hat{\gamma},X_S,X_B)}-\beta)||_2^2\leq \inf_W n_S^{-1}E||X_S(\hat{\beta}_{W}-\beta)||_2^2+O(n_S^{-1})
\end{eqnarray*}

\textbf{Theorem 2:} The higher order optimal weight $W_h(\beta,\gamma,X_S,X_B)$ and its estimated version $\hat{\beta}_{\hat{W}_h(\hat{\beta},\hat{\gamma},X_S,X_B)}$ approximately minimize $E||\hat{\beta}_W-\beta||_2^2$  at $o(n_S^{-2})$ level in the sense that 
\begin{eqnarray*}
&E||\hat{\beta}_{W_h(\beta,\gamma,X_S,X_B)}||_2^2\leq \inf_W E||\hat{\beta}_{W}-\beta||_2^2+o(n_S^{-2})\\
&E||\hat{\beta}_{\hat{W}_h(\hat{\beta},\hat{\gamma},X_S,X_B)}||_2^2\leq \inf_W E||\hat{\beta}_{W}-\beta||_2^2+O(n_S^{-2})
\end{eqnarray*}

\textbf{Theorem 3:} Assuming $n_B/n_S\rightarrow r\in (0,\infty)$ and $n_S\rightarrow \infty$, the weighted estimator based on estimated higher order optimal weights is more efficiency than using small data only, i.e., $E||\hat{\beta}_{\hat{W}_h(\hat{\beta},\hat{\gamma},X_S,X_B)}-\beta||_2^2
\leq E||\hat{\beta}_{I}-\beta||_2^2$ hold asymptotically when $n_S\rightarrow \infty$.\\


Here we provide the general idea for the proof while the details can be found in the Appendix. We can use the second or high order expansion of $\hat{\beta}$ to obtain the optimal weight as a function of $\beta$ and $\gamma$ and the improvement of using weighted matrix is of the order of $O(n_S^{-2})$ in comparison to using small data only. Then we showed that the difference between MSE based on these estimated optimal weight and oracle optimal weight are positive with approximation error in the order of $o(n_S^{-2})$. The details of the proof of these theorems could be found in appendix.



\section{Analysis of ACC data} 
The Asia Cohort Consortium (ACC) is a collaborative effort born out of the need to study Asian population, seeking to understand the relationship between genetics, environmental exposures, and the etiology of a disease through the establishment of a cohort of at least one million healthy people around different countries in Asia, followed over time to various disease endpoints and death. This pooling project, with its huge sample size across 29 subcohorts from 10 Asian countries (\url{https://www.asiacohort.org/ParticipatingCohorts/index.html}), provides the perfect opportunity to explore informative relationships (association of exposure with disease, genome variability with disease etc) among major Asian ethnic groups. 

Over the last few decades, obesity has become an important health issue in many countries. According to World Health Organization estimates, more than a billion adults around the world are overweight, and at least 300 million of them are obese \citep[see][]{Abelson2004}. Many epidemiological studies have found association between the body-mass index (BMI) and a variety of health outcomes, including mortality \citep[see][]{Haslam2005}. However, most of these inferences have been drawn from studies in populations of European origins, and very little focus has been given to the relationship between BMI and the overall risk of death among Asians, who account for more than 60\% of the world population \citep[see][]{Zheng2011}. The data collected as part of the ACC can be used to answer these important questions.

To show the usefulness of our proposed methodology in a practical setting, we use data from the ACC to explore the relationship between BMI and mortality. In particular, we concentrate only on the cohorts from China - data from the Shanghai Cohort Study (SCS) is used to form our small data, while data from rest of the Chinese subcohorts - China Hypertension Survey Epidemiology Follow-up Study (CHEFS), Linxian General Population Trial Cohort, Shanghai Men's Health Study (SMHS) and Shanghai Women's Health Study (SWHS) are pooled together to form the initial big data. Since the SCS cohort only included males, we decide to restrict the big data to include only male participants from the other subcohorts (which completely excluded the SWHS). For individuals in the small data, enrollment started in 1986 and the study is continued till 2007, while for the pooled large data, enrollment started in 1985, and the last year of follow-up is 2011. Missingness in covariates is not a big concern (no missingness in the small data, and only 0.79\% missingness in the large data). The baseline age distribution of the individuals is found to be different in the small and the large data, and since mortality is a definite function of age, for better comparability, we decide to restrict the two datasets such that they contain individuals whose baseline age varied between 50 and 60. And because methods described in this paper pertain to binary outcomes only, and time to follow-up varies for different individuals in the two datasets, we decide to only consider the first year of follow-up for each individual. Firstly, this makes the binary statuses of mortality comparable for individuals in the two datasets, and secondly the short period of follow-up ensures that we do not lose too many individuals who are lost to follow-up. Such individuals forms only 0.06\% of the small data and 2.88\% of the large data, and are removed from the analysis. After performing all these data management steps, the small data is found to contain 10675 individuals with 40 mortality events, while the large data is found to contain 46779 individuals with 206 events. Apart from BMI, baseline age is also included as a covariate in the model, as well as indicators for each individual's smoking and drinking habits, as these covariates have been proven to be important predictors of mortality in many settings. 


We start off with analyzing the small and the big data separately first, using the standard logistic regression model, and then by pooling them together. We then estimate the regression coefficients using the proposed weighted shrinkage methods, namely, with the optimal second order weights ($W_2$), the optimal higher order Edgeworth weights ($W_{h}$), and for comparison, the optimal James-Stein weights (JSP). We also obtained the penalized estimates, using the $L_1$ and $L_2$ procedures. 

The estimates and their standard errors for the various procedures are presented below in Table \ref{tab:realdat}. As can be seen, the pooled procedure obtains the lowest standard errors, owing to the fact that it uses the entirety of the big and the small data, but it also means that the estimates for this procedure are inherently biased towards the ones that we obtain from the big data itself, as it contains a lot more information (than the small data) because of its size, so naive pooling inappropriately shifts most of the focus to the big data itself. The weighted shrinkage procedures seem to be better adjusted in this respect, with estimates shrunk somewhat but much closer to the ones that we see from the small data itself, but with much lower standard errors than the small data estimates. The optimal second order and higher order Edgeworth weights ($W_2$ and $W_{h}$) perform similarly in this regard, and has lower standard errors than the estimates from the James Stein adjusted weights JSP, except for BMI, in which case the JSP procedure obtains a lower standard error than $W_2$ or $W_h$, however the estimate for BMI obtained by JSP is shrunk completely to that obtained from the big data itself. Among the penalized procedures, $L_1$ seem to borrow more strength from the big data, and thus has lower standard errors and higher amount of shrinkage, while the estimates for the $L_2$ procedure seem to be closer to the small data estimates, and thus have higher standard errors.


\begin{table}[ht!]
\setlength{\tabcolsep}{2pt}
\centering
\begin{tabular}{ll c@{\hskip 0.2in}c@{\hskip 0.2in}c@{\hskip 0.2in}c@{\hskip 0.2in}c@{\hskip 0.2in}c@{\hskip 0.2in}c@{\hskip 0.2in}c}
\hline
&& \textbf{Small} & \textbf{Big} & \textbf{Pool} & \textbf{$W_2$}& \textbf{$W_{h}$} & \textbf{JSP}&\textbf{$L_1$} & \textbf{$L_2$} \\
\hline
\textbf{(Intercept)} &\textbf{Est}& -4.80 & -9.89 & -9.08 & -5.25 & -5.26 & -6.27 &-9.04 &-4.85  \\
& \textbf{Std Err} & 3.21 & 1.48 & 1.34 & 1.88 & 1.88  &2.32& 1.42 & 1.77 \\
\textbf{Age at Baseline}&\textbf{Est} & 0.03 & 0.12 &0.10& 0.04& 0.04 &0.05& 0.10 &0.03  \\
& \textbf{Std Err} & 0.05 & 0.02 & 0.02 & 0.03 & 0.03 &0.04& 0.02 & 0.04 \\
\textbf{Body Mass Index}&\textbf{Est} & -0.14 &-0.09& -0.09& -0.13& -0.13& -0.09& -0.12& -0.14\\
& \textbf{Std Err} & 0.06 & 0.02 & 0.02 & 0.03 & 0.03  &0.02& 0.03 & 0.05 \\
\textbf{Ever Smoked}&\textbf{Est} & 0.71& 0.08& 0.20& 0.65& 0.65&  0.55& 0.62& 0.71 \\
& \textbf{Std Err} & 0.38 & 0.16 & 0.14 & 0.22 & 0.22 & 0.29 & 0.30 & 0.40 \\
\textbf{Ever Used Alcohol}&\textbf{Est} & 0.06 &-0.37& -0.30& 0.02& 0.02 & -0.13& -0.02& 0.06 \\
& \textbf{Std Err} & 0.33 & 0.16 & 0.14 & 0.18 & 0.18  &0.20& 0.27 & 0.31 \\
\hline
\end{tabular}
\vspace{10pt}
\begin{center}
\caption{Estimates and their standard errors from ACC data analysis %
    \label{tab:realdat}}
\end{center}
\end{table}

\section{Discussion} 
In this paper, we proposed better estimators that allow more accurate estimation of the regression coefficient and the risk prediction for our target population using information from another different population with more observations. Although the expansion and detailed form of the weight we provided are specifically for logistic regression, the optimal weight formula is general in terms of the expansion formula $C$, $D$, $E$s. So the framework we proposed here could be extended to generalized linear model and estimating equation models straightforward thought the more complicated computation of Edgeworth expansion for these estimating equation based estimator need to be derived for the optimal estimation weight. 

To utilize the big data, although we do not need to know the exact relationship between the small and big data in terms of association strength, we do need the model form in big data to be correctly specified. In our setting, the same logistic form need to hold for both the big and the small data. When the covariate is limited and is categorical, this assumption is weak and easy to satisfy. When there is continuous covariate, we can apply existing model checking tools to the big data to check whether our model assumption holds.

In our analysis of the ACC data, we only concentrated on the first year of follow-up, because the methods presented in this paper are relevant only for binary outcomes, and the short period of follow-up ensured that we did not lose too many individuals to loss to follow-up, which would have otherwise introduced unforeseen sources of bias in our analysis. However in doing so, we lost a lot of rich information which is contained in the time to follow-up data. This shows the need to extend our methods to the case when we have time to event data, and this indeed is one of our future research goals.

\section{Supplementary Materials}
Web Appendices referenced in Sections 2 and 3 are available with this article at the Biometrics website on Wiley Online Library.

\label{lastpage}
\end{document}